\documentclass[journal=nalefd,manuscript=communication]{achemso}
\usepackage{stmaryrd}
\usepackage{amsmath,amssymb}
\usepackage{soul}
\usepackage{graphicx}
\usepackage{bm}
\usepackage{graphics}
\usepackage{xfrac}
\usepackage[usenames,dvipsnames]{xcolor}
\usepackage[normalem]{ulem}

\usepackage[version=3]{mhchem} 

\author{Le Qiao}
\email{leqiao.lq@gmail.com}
\affiliation[Uottawa University]
{Physics Department, University of Ottawa, Ottawa, Ontario, K1N 6N5, Canada}
\author{Kai Szuttor}
\affiliation[Stuttgart University]
{Institute for Computational Physics, University of Stuttgart, Stuttgart, D-70569, Germany}
\author{Christian Holm}
\affiliation[Stuttgart University]
{Institute for Computational Physics, University of Stuttgart, Stuttgart, D-70569, Germany}
\author{Gary W. Slater}
\email{gary.slater@uOttawa.ca}
\affiliation[Uottawa University]
{Physics Department, University of Ottawa, Ottawa, Ontario, K1N 6N5, Canada}

\title{Ratcheting charged polymers through symmetric nanopores using pulsed fields: Designing a low pass filter for concentrating polyelectrolytes}

\keywords{nanopore, DNA, translocation, separation, ratchet, AC}

\begin{document}


\begin{abstract}
We present a new concept for the separation of DNA molecules by contour length that combines a nanofluidic ratchet, nanopore translocation and pulsed fields. Using Langevin Dynamics simulations, we show that it is possible to design pulsed field sequences to ratchet captured semiflexible molecules in such a way that only short chains successfully translocate, effectively transforming the nanopore process into a low pass molecular filter. We also show that asymmetric pulses can significantly enhance the device efficiency. The process itself can be performed with many pores in parallel, and it should be possible to integrate it directly into nanopore sequencing devices, increasing its potential utility.
\end{abstract}
Size-dependent separation of DNA molecules is crucial for molecular analyses such as genome sequencing. To date, gel electrophoresis is the most commonly used separation method in the laboratory. However, the gel-based method has many disadvantages, such as difficult sample recovery, low separation efficiency, and poor separation resolution for long DNA molecules.\cite{dorfmanGelElectrophoresisMicrofluidic2013}. Alternatively, "lab-on-chip" micro/nanofluidic devices show unique advantages in size-based DNA separation as they are relatively inexpensive,  and sometimes allow label-free continuous separation\cite{sonkerSeparationPhenomenaTailored2019,cabodiContinuousSeparationBiomolecules2002a}. Different mechanisms including dielectrophoresis  \cite{dorfmanDNAElectrophoresisMicrofabricated2010a}, ratchet rectification \cite{desruisseauxTrappingElectrophoresisRatchets1998}, entropic trapping \cite{streekTwostateMigrationDNA2005}, Ogston sieving \cite{slaterExactlySolvableOgston1996a}, confinement-driven separation, or a combination of those, have been used to achieve the goal of separation. Among them, many of the nanofluidic devices that have been proposed rely on generating mobility differences using a geometric constraint. In such designs, all species in the sample can only move in the same direction, so the separated fractions must be collected at different times at the end of the channel. In addition, all analytes pass through the same space at rather high speeds \cite{cabodiContinuousSeparationBiomolecules2002a}, making continuous separation extremely difficult.

\begin{figure}[htbp!]
\begin{center}
\includegraphics[scale=1.5]{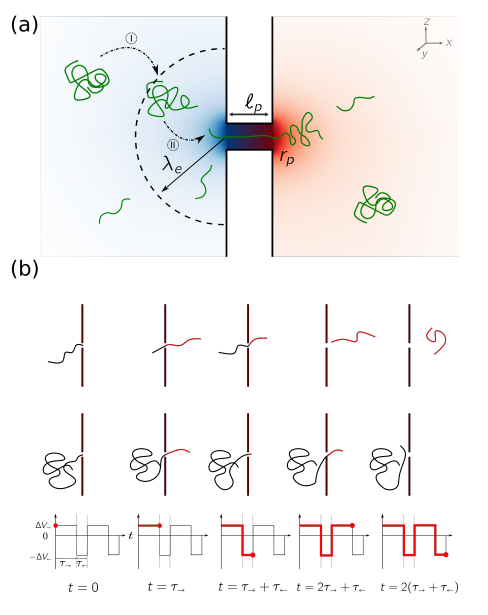}
\end{center}
\caption{(a) A schematic view of polymer translocation through a nanopore driven by an external voltage; the background codes for the strength of the electric field (more opacity meaning higher fields). The curved arrows identify two stages that are part of the capture process: I Analytes randomly diffuse towards the capture radius; II Field-driven drift towards the nanopore. The nanochannel is of radius $r_p$ and length $\ell_p$. The dashed line depicts the nominal capture radius ($\lambda_e$) separating the diffusion-dominated and drift-dominated regions. (b) The protocol for bidirectional separation of two different sized molecules using ZIFE pulses; the pulses are asymmetric but remain unbiased when $\kappa \equiv \frac{\tau_\shortrightarrow}{\tau_\shortleftarrow}=\frac{\Delta V_\shortleftarrow}{\Delta V_\shortrightarrow}\ne 1$.}
\label{Fig:system}
\end{figure}

Polymer translocation is the process by which a polymeric chain is forced to move through a wall via a nanopore or nanochannel \cite{mellerVoltageDrivenDNATranslocations2001,wanunuNanoporesJourneyDNA2012,palyulinPolymerTranslocationFirst2014,shiNanoporeSensing2017,buyukdagliTheoreticalModelingPolymer2019,xueSolidstateNanoporeSensors2020,wenFundamentalsPotentialsSolidstate2021}. Typically, an electric field is used to drive this process for charged molecules like DNA and other polyelectrolytes; the field is then used to both, capture the molecules, and force translocation (Fig.~\ref{Fig:system}a). Several low-resolution devices have been built using this general idea \cite{xueSolidstateNanoporeSensors2020,wenFundamentalsPotentialsSolidstate2021}. In theory, one can favour the translocation of the smallest molecule in a mixture by applying the translocating field (\textit{i.e.}, once the molecule is captured) during a forward time $\tau_\shortrightarrow$ comparable to its mean DC translocation time $\overline{\tau}$, after which the field polarity would be reversed for a duration $\tau_\shortleftarrow$ long enough to disengage all molecules that have failed to translocate. Gupta \textit{et al.} have proposed such an idea to separate $\lambda$ DNA and $T4$ DNA molecules, which have very different molecular sizes, in long nanochannels\cite{guptaDNATranslocationShort2014}. However, the distribution of translocation times is very broad \cite{wanunuNanoporesJourneyDNA2012}: if the molecules are close in size, these distributions will significantly overlap and $\tau_\shortrightarrow$ will have to be chosen very short ($\tau_\shortrightarrow \ll \overline{\tau}$) to inhibit the translocation of the longer chains -- with an obvious unfavourable effect on translocation rates (as usual, the conditions for high purity and high numbers compete with each other).  

Recently, Zhan \textit{et al.}\cite{zhanDetectionSeparationSingleStranded2021a} have shown that it is possible to obtain DNA size-dependent mobilities by tuning the nanopore size to make the translocation favour certain chain lengths. This is due to the fact that the electric force driving the capture outside the pore, the entropic barrier to enter the nanopore, as well as the friction inside the pore, all depend on the radius/length ratio of the nanochannel. Instead of changing the nanochannel size, here we propose to use pulsed electric fields to selectively ratchet \cite{chialvoAsymmetricUnbiasedFluctuations1995,baderDNATransportMicromachined1999,wangNetMotionCharged2013,mondalRatchetRectificationEffect2016, parkSilicoConstructionFlexibilitybased2019} captured molecules across a symmetric nanochannel. In principle, the fact that we have different molecular conformations on both sides of the wall during translocation can be used to drive a ratchet process even with a perfectly symmetric nanochannel. We also show that asymmetric pulsed fields improve ratcheting. More precisely, we use the ratchet effect to move small molecules in one direction (translocation) while moving the larger ones in the opposite direction (retraction), thus enabling bidirectional separation (see Fig.~\ref{Fig:system}b). Importantly, it is possible to control the critical molecular size by tuning the frequency of the electric field. We explore two versions (or protocols): one in which the ratchet effect starts upon capture, while the second one combines capture and translocation.
Both computer simulation data and the tension propagation theory \cite{sakaueNonequilibriumDynamicsPolymer2007,sakaueSuckingGenesPores2010,saitoDynamicalDiagramScaling2011,sarabadaniDrivenTranslocationSemiflexible2017,ikonenUnifyingModelDriven2012} show that the mean translocation time of a relaxed chain of $N$ monomers under a constant driving force applied to the monomers inside the nanochannel scales like $\overline{\tau}\sim N^\alpha$, where $\alpha$ increases with $N$ to reach the asymptotic value $\alpha=1+\nu$ ($\nu=3/5$ is Flory's exponent). The capture time $\tau_c$ of a chain, however, is independent of $N$ \cite{wanunuElectrostaticFocusingUnlabelled2010}. Since $\tau_c \gg \overline{\tau}$, translocating a solution of DNA molecules using a DC field is not expected to change the relative abundance of the different molecular sizes. This is the problem we are proposing to address using ratcheting pulsed fields.

Two elements are required to build a ratchet that can rectify otherwise random motion: keeping the system out of equilibrium and symmetry breaking. The first condition can be achieved using either a stochastic or a deterministic external perturbation \cite{magnascoForcedThermalRatchets1993,gulyaevNanotransportControlledMeans2020, kenwardPolymerDeformationBrownian2008}, such as a pulsed field. The symmetry can be spatially \cite{kenwardPolymerDeformationBrownian2008,slaterBidirectionalTransportPolyelectrolytes1997,hammondDifferentialTransportDNA2000,malgarettiConfinedBrownianRatchets2013,heidariMechanismRectificationPolymer2020} and/or temporally broken \cite{slaterBidirectionalTransportPolyelectrolytes1997,tessierStrategiesSeparationPolyelectrolytes2002}. In this paper, some spatial asymmetry comes from the conformational and size differences between the polymer segments on each side of the wall, while temporal asymmetry comes from pulsed fields such as the Zero-Integrated-Field Electrophoresis (ZIFE) pulses shown in Fig.~\ref{Fig:system}b and previously used in gel electrophoresis \cite{brassardPulsedFieldSequencing1992}.

Our Langevin Dynamics simulations use coarse-grained bead-spring chains with the bead-bead and bead-wall excluded volume interactions modeled using the purely repulsive Weeks-Chandler-Andersen (WCA) potential \cite{weeksRoleRepulsiveForces1971}
\begin{equation}
U_\mathrm{WCA} (r) = \label{EQ:WCA}
\begin{cases}
4 \epsilon \left[ \left( \frac{\sigma}{r}\right)^{12} - \left( \frac{\sigma}{r} \right)^6 \right] +\epsilon   &\text{for } r < r_c  \\
0 &\text{for } r \geq r_c.
\end{cases}
\end{equation}
The nominal bead size $\sigma$ is used as the unit of length and the well depth $\epsilon=k_BT$ as the unit of energy, while the cutoff length $r_c \! = \! 2^{1/6}\,\sigma$ makes $U_{WCA}$ purely repulsive. The Finitely-Extensible-Nonlinear-Elastic (FENE) potential $U_{\textrm{FENE}}(r) = - \tfrac{1}{2} K_{\textrm{FENE}}~ r^2_0 ~ \textrm{ln} \left( 1 - \sfrac{r^2}{r_0^2} \right)$ is used to connect adjacent monomers \cite{grestMolecularDynamicsSimulation1986}; the maximum extension is $r_0 \!= \! 1.5\,\sigma$ and the spring constant is $K_{\textrm{FENE}} \! = \! 30\,\epsilon / \sigma^2$. The combined energy $U_{WCA}+U_{FENE}$ gives a mean bond length of $\langle b\rangle\approx0.96~\sigma$. We control the polymer stiffness via the angular harmonic potential $U_{\textrm{Bend}}(\theta) = \tfrac{1}{2}K_{\textrm{Bend}} \left(\theta- \pi\right)^2$, where $\theta$ is the angle between two consecutive bonds and $K_{\textrm{Bend}}=5\,\epsilon$; the resulting free solution polymer persistence length is $L_p \approx 5\,\sigma$. 

The electric potential outside the pore is given by \cite{farahpourChainDeformationTranslocation2013}
\begin{equation}
 \label{seq:sol}
  V(\zeta,\eta,\phi) = \Delta V~ \tfrac{r_e }{r_p}~\arctan \left[ \sinh(\zeta) \right],
\end{equation}
where $\Delta V$ is the total potential difference across the device,  $r_e=r_p/(\frac{2\ell_p}{r_p}+\pi)$ is the characteristic length of the electrostatic potential outside a nanochannel of radius $r_p$ and length $\ell_p$, while $\zeta \! \in \! (- \infty, + \infty)$, $\eta \! \in \! [ 0, \pi ]$ and $\phi \! \in \! [ 0, 2 \pi ]$ are the oblate spherical coordinates. The potential drop across the channel is $\delta V =\Delta V \times {2\ell_p r_e}/{r_p^2}$, corresponding to an electric field $E_p \! = \! {\delta V}/{\ell_p} \! = \! \Delta V \times {2r_e}/{r_p^2}$. We use $r_p \! = \! 1.5\, \sigma$ (which ensures single file dynamics) and $\ell_p \! = \! 5\, \sigma$; the potential drop between an electrode ($r \! \to \! \infty$) and the pore is then $\approx \! \frac{1}{6}\,\Delta V$ while the drop inside the channel is $\delta V \! \approx \! \frac{2}{3} \,\Delta V$. The nominal capture radius\cite{qiaoVoltagedrivenTranslocationDefining2019,qiaoCaptureRodlikeMolecules2020} for a chain of $N$ monomers (each with a charge $Q$), $\lambda_e={N\Delta VQr_e}/{k_BT}$, will be used as a field intensity.

The LD equation of motion for a monomer is
\begin{equation}
m \dot{\vec{v}} = \vec{\nabla} U (\vec{r}) - \xi \vec{v} + \sqrt{ 2 \xi k_\mathrm{B} T }~ \vec{R}(t),
\label{Eq_lD}
\end{equation}
where $\vec{\nabla} U(\vec{r})=\vec{\nabla}(U_\mathrm{WCA}+U_\mathrm{FENE}+U_\mathrm{Bend}+U_\mathrm{E})$ is the sum of the conservative forces, $U_\mathrm{E} (\Vec{r})$ is the electric potential, $m$ and $\xi$  are the mass and friction coefficient of a monomer, and $-\xi \vec{v}$ is the damping force due to the fluid. The last term on the \textit{rhs} is the uncorrelated noise that models random kicks from the solvent. The random variable $\vec{R}(t)$ satisfies $\langle R_i(t) \rangle = 0$ and $\langle R_i(0) R_j(t) \rangle = \delta(t) \delta_{ij}$, where $\delta (t)$ is the Dirac delta function and $i,j \in [x,y,z]$. The unit of time $\tau_o={\sigma^2\xi}/{k_BT}$ is chosen to be the time needed for a monomer to diffuse over its own size $\sigma$. In our simulations the integration time step is $\Delta t= 0.01\,\tau_o$. The systems are in a box with a minimum size of $1.5\,\lambda_e$ (as measured from the nanopore) and periodic boundary conditions are applied. 

Although we use dimensionless variables, it is useful to look at potential examples. Since dsDNA has a persistence length $L_p \! \approx \! 50\,nm$, our parameters then correspond to $\sigma \! = \! 10\,nm$, $r_p \! = \! 15\, nm$ and $\ell_p \!= \!50\,nm$.
For ssDNA, these numbers would be $L_p \approx 2\,nm$, giving $\sigma=0.4\,nm$, $r_p=0.6\,nm$ and $\ell_p=2\,nm$. 

Feedback control of the applied voltage by sensing the nanopore current has been used previously for both nanopore translocation\cite{lathropMonitoringEscapeDNA2010,rahmanDemandDeliveryAnalysis2019} and nanopore fabrication\cite{waughSolidstateNanoporeFabrication2020}. Our first protocol sorts polymers by size from a binary mixture using a feedback mechanism. In short, it employs a DC field to capture polymers and a current detector to determine when a chain end has entered (or left) the nanopore. We start each capture simulation by placing the center of mass of a polymer with a random equilibrium
conformation (radius of gyration $R_{go}$) at a radial distance $r_o \! = \! \lambda_e \! \approx \! 5 \! - \! 10\,R_{go}$ right above the nanopore. When one end monomer has reached the channel's mid-point, the DC field is replaced by the pulsed field (Fig.~\ref{Fig:system}b) to selectively translocate the shortest chains. Details on polymer capture are described in Supporting Information. 

\begin{figure}[htbp!]
\begin{center}
\includegraphics[scale=1.5]{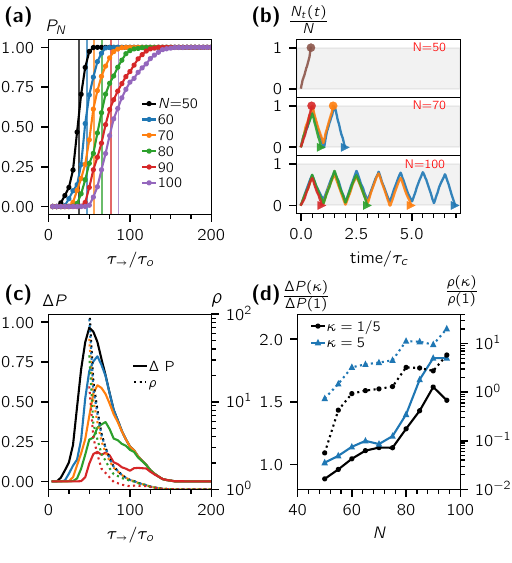}
\end{center}
\caption{Pulsed field data for captured polymers of size $N\!=\!50\!-\!100$ (ensemble size $\Omega\,=\,1000$). (a) Translocation probability $P_N$  \textit{vs} forward pulse duration $\tau_\shortrightarrow$ for unbiased pulses ($\kappa=1$) with a field intensity $\lambda_e(N)=\frac{N}{50}\times168.9\,\sigma$. The vertical lines give the mean DC translocation times. (b) Fractional number of monomers $N_t(t)/N$ that have passed the mid-point of the channel \textit{vs} time for three chain lengths ($\kappa=1)$. The duration of one complete pulse cycle, $\tau_c=\tau_{\rightarrow}+\tau_{\leftarrow}=2\tau_{\rightarrow}=100\,\tau_o$, is used to rescale the time axis. The filled circles show the end of a successful translocation while the triangles mark full retraction.
(c) Same data plotted to show the differential probability $\Delta P(N,100)$ (solid lines; left y-axis) and the probability ratio $\rho(N,100)$ (dotted lines; right y-axis).
(d) Ratios $\Delta P(N,100,\kappa)/\Delta P(N,100,1)$ (solid lines; left y-axis) and $\rho(N,100,\kappa)/\rho(N,100,1)$ (dotted lines; right y-axis) \textit{vs} polymer size $N$ for $\kappa=5$ and $\kappa=1/5$. The low and high field intensities are $\lambda_e(N)=\frac{N}{50}\times56.3\,\sigma$ and $\lambda_e(N)=\frac{N}{50}\times281.5\,\sigma$, respectively.}
\label{fig:ZTP}
\end{figure}

Figure \ref{fig:ZTP}a shows the probability of translocation $P_N$ as a function of the pulse duration $\tau_\shortrightarrow$ when a symmetric pulsed field ($\kappa \equiv \frac{\tau_\shortrightarrow}{\tau_\shortleftarrow}=\frac{\Delta V_\shortleftarrow}{\Delta V_\shortrightarrow} \! = \! 1$) is used. The vertical lines indicate the mean DC translocation times $\overline{\tau} \sim N^{1.21}$. Not surprisingly, we observe sigmoidal curves with midpoints located roughly at $\tau_\shortrightarrow \approx \overline{\tau}(N)$ and widths directly related to that of the distribution function of DC translocation times. Some trajectories are shown in Fig.~\ref{fig:ZTP}b for $\tau_\shortrightarrow=50\,\tau_o$: the short $N=50$ chains translocate before the first polarity reversal while the $N=70$ chains oscillate back and forth for a few cycles ($\approx 50\%$ of them ultimately succeed). The $N=100$ chains do not translocate: they simply retract after a few cycles. 

An ideal separation device would have both a high differential efficiency $\Delta P(N_1,N_2) = P_{N_1}-P_{N_2}$ and an excellent purity ratio $\rho(N_1,N_2) = P_{N_1}/P_{N_2}$. Figure~\ref{fig:ZTP}c shows how $\Delta P(N,100)$ and $\rho(N,100)$ vary as a function of the pulse duration $\tau_\shortrightarrow$ for a field intensity $\lambda_e(N)=\frac{N}{50}\times168.9\,\sigma$ and symmetric pulses ($\kappa=1$). As expected, both parameters increase with the molecular size difference, with $\Delta P(N,100)$ increasing from 0.2 (for $N=90$) to 0.95 (for $N=50$), while the corresponding ratio $\rho(N,100)$ increases from 3 to 80. The ratio $\rho$ diverges at short pulse durations because the $N=100$ chains then fail to translocate; however, $\Delta P$ then rapidly decays. The best overall performance is thus located near the maximum of $\Delta P(N,100)$. The performance is rather poor when $(N_1-N_2) \ll N_2$, as expected.

Since conformational asymmetry alone does not produce a strong ratchet effect, we now explore the impact of adding a pulse asymmetry $\kappa \ne 1$. We explore two different cases: $\kappa<1$ (the translocation is driven by the higher field intensity) and $\kappa>1$ (the low field drives the translocation). For a fixed asymmetry $\kappa$, both $\Delta P(N,100)$ and $\rho(N,100)$ are enhanced compared to the $\kappa=1$ case -- see Fig.~\ref{fig:ZTP}d, with the $\kappa>1$ ZIFE pulses providing better results. We believe that both the conformational entropic barrier due to the pore confinement\cite{muthukumarTheoryCaptureRate2010} and the non-monotonic monomer translocation rate during the translocation process play important roles here. During forward pulse duration $\tau_\shortrightarrow$, the low field is sufficient for shorter chains to cross the energy barrier and the long $\tau_\shortrightarrow$ ensures them to approach the fast translocation regime (the post-propagation stage\cite{saitoDynamicalDiagramScaling2011,sarabadaniDrivenTranslocationSemiflexible2017}); during the backward pulse duration $\tau_\shortleftarrow$, the high field but short $\tau_\shortleftarrow$ can rapidly retract the long polymer from nanopore as a smaller portion of the chain translocated in previous $\tau_\shortrightarrow$ as compared to shorter chains. In fact, the ratio $\rho(95,100)$ is increased by a remarkable factor of $\approx 20$ compared to symmetric pulses. While for polymer chains with large molecular size differences (e.g., $N=50$ and $N=100$) the separation is already excellent with symmetric pulses, the ratio $\rho(50,100)$ actually drops when the pulsed field becomes asymmetric. In particular, the $\kappa=5$ pulses lead to a significant decrease; this is due to the fact that high forward field intensities actually help the $N=100$ chains to overcome the conformational entropic barrier near the pore entrance.

However, the feedback approach presented above does not make it possible to use multiple nanopores in parallel (the feedback loop is necessary because unbiased ZIFE pulses cannot capture polymers even when $\kappa \ne 1$). To avoid this issue, we now bias the pulses by increasing the forward time from $\tau_\shortrightarrow$ to $\tau_f \! = \! \tau_{\shortrightarrow}+\tau_{\Rightarrow}$. The dimensionless mean field intensity $\Phi(\tau_{\Rightarrow},\kappa,\tau_{\shortrightarrow}) \! = \! \frac{\langle \Delta V \rangle}{\Delta V_{\shortrightarrow}} \! = \! \frac{\tau_{\Rightarrow}}{\tau_{\Rightarrow}+\tau_{\shortrightarrow}(1+1/\kappa)}$ will be used as a measure of the bias. 

Before translocation, the field-driven deterministic time to move from position $r_o$ to $r \! < \! r_o$ under the action of the electric is $\tau_E(r_o,r) \! = \! \int_{r_o}^{r} {v_d^{-1}(r^{\prime})} {\mathrm{d}r^{\prime}} \! = \! (r_o^3-r^3)/3\lambda_e D$ \cite{qiaoVoltagedrivenTranslocationDefining2019}, with a similar expression for reverse pulses (here $D$ is the center-of-mass diffusion coefficient of the polymer chain). The position after $M$ complete cycles is then $r(r_o,M)=\sqrt[3]{r_o^3-3M\lambda_eD\tau_{\Rightarrow}}$ (note that unbiased $\tau_{\Rightarrow} \! = \! 0$ pulses lead to $r \! = \! r_o$ or no net motion towards the pore). The largest value of $M$ ($ \in \mathbb{N}_0$) for which the argument of the cubic-root is positive gives the number $M_o$ of complete pulse cycles needed to capture a polymer initially located at $r \! = \! r_o$. The final step, during which the chain ends try to find the pore entrance, thus starts with a forward pulse of reduced duration $\tau_f^o=\tau_f-\tau_E(r(r_o,M_o),0)=\tau_{\shortrightarrow}+\tau_{\Rightarrow}-r(r_o,M_o)^3/3\lambda_e D$. We tested that $\tau_f^o \in [0,\tau_f]$ is essentially a random number when we vary $r_o$, even though this is a deterministic result for a particle (data not shown). Since both Brownian motion and the time required by the polymer ends to find the nanopore add randomness, we test the biased ratchet process by averaging over a flat distribution of initial pulse times $\tau_f^o$. 

\begin{figure}[htbp!]
\begin{center}
\includegraphics[scale=1.5]{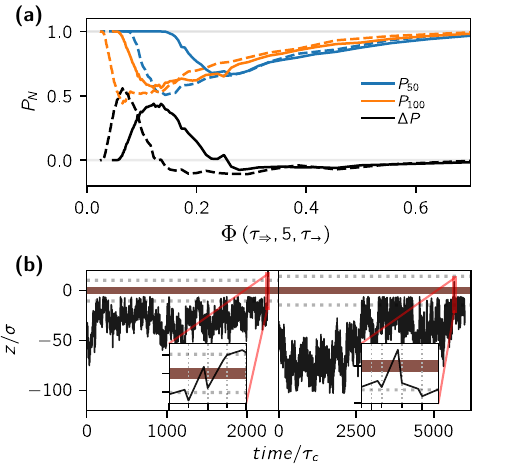}
\end{center}
\caption{(a) Translocation probability $P_N$ \textit{vs} dimensionless mean field intensity $\Phi(\tau_\Rightarrow,\kappa=5,\tau_\shortrightarrow)$. The time biases are $\tau_{\Rightarrow}=15$ (dashed lines) and $30\,\tau_o$ (solid lines); the mean field is changed by varying $\tau_{\shortrightarrow}$ between 25 and 500$\,\tau_o$. (b) Trajectories of the end monomer of chains of size $N=50$ (left panel) and $100$ (right panel). The bias is $\tau_{\Rightarrow}=30 \tau_o$ and the total cycle duration is $\tau_c=\tau_{\shortrightarrow}+\tau_{\Rightarrow}+\tau_{\shortleftarrow}=210\,\tau_o$, which gives a mean field intensity $\Phi(30,5,150)=\frac{1}{7}$. The y-axis gives the monomer's vertical distance to the wall center. The gray dashed lines mark the distance $R_{go}$ from the wall surface, with of $R_{go}=7.9$ and $12\,\sigma$ for $N=50$  and $100$, respectively}
\label{fig:biasedAC}
\end{figure}

Figure~\ref{fig:biasedAC}a shows the probabilities $P_{100}$, $P_{50}$ and $\Delta P=P_{50}-P_{100}$ as a function of the mean field intensity $\Phi$ for pulse biases $\tau_{\Rightarrow} \! = \! 30\, \tau_o$ and $15\,\tau_o$ (both values are smaller than the chains' mean DC translocation times $\overline{\tau}_{50}=111.6\,\tau_o$ and $\overline{\tau}_{100}=257\,\tau_o$). Since we saw in Fig.~\ref{fig:ZTP}d that $\kappa \! = \! 5$ provides an excellent separation ratio, here we fix $\kappa$ and use $\tau_\shortrightarrow$ to tune the mean field $\Phi(\tau_{\Rightarrow},5,\tau_{\shortrightarrow})$. In the short pulse regime $(\tau_{\shortrightarrow}+\tau_{\shortleftarrow}) = (1+1/\kappa) \tau_{\shortrightarrow} \ll \tau_{\Rightarrow}$, the AC component is negligible and we basically have a DC field of intensity $\Phi=1$: both molecules then easily translocate ($P_N \to 1$). In the opposite long pulse limit $\tau_\shortrightarrow > \overline{\tau}_N > \tau_\Rightarrow$, a chain of size $N$ translocates immediately during the first part of the first cycle and we again have $P_N \to 1$ although $\Phi$ is now small. Note that since this is happening at a molecular size dependent critical value of the pulse duration $\tau_\shortrightarrow(N)$, there is a range where separation is possible, as shown. As expected, the maximum value of $\Delta P=P_{50}-P_{100}$ increases when the bias is reduced (from $\tau_{\Rightarrow}=30 \tau_o$ to $15 \tau_o$ here) because we are moving towards the conditions present in Fig.~\ref{fig:ZTP}d. In other words, the bias reduces $\Delta P$ because it increases the probability of translocation for both molecular sizes. 

However, the bias allows the pulses to also capture the polymer chains here: Fig.~\ref{fig:biasedAC}b shows two capture-and-translocation trajectories with $\Phi(30,5,150)=0.14$. The $N=100$ and $50$ chains start from $r_o=50\,\sigma$ right above the nanopore with a reflecting wall at $z=100\,\sigma$. Similar to DC capture, both chains spend a considerable amount of time before one end of the chain enters the nanopore. The $N=50$ chain translocates within 2 cycles after capture while the $N=100$ chain quickly retracts. In short, a bias added to the ratcheting pulses allows us to both capture and translocate molecules; the size specificity is maintained but reduced. On the other hand, the absence of a feedback loop means that one can run hundreds or thousands of pores in parallel (or/and in series), a major advantage in some cases. Optimization of parameters for specific applications is beyond the scope of this letter.

The combination of molecular size-independent capture rates and broad distributions of translocation times greatly limits the use of nanopore translocation as a process that can modify the concentration ratio of DNA solutions. We have previously proposed an experimental approach for small (point-like) particles that have different diffusion coefficients but the same electrophoretic mobility: this scheme made use of \textit{on/off} pulsed-fields \cite{qiaoEfficientKineticMonte2021} to modify their capture rates. 

In this Letter, we now introduce the basic features of a low pass molecular filter device for molecules such as dsDNA. The key idea is to use pulsed electric fields to ratchet captured polymer chains through the nanochannel, and to design the pulses to favour the translocation of small molecules. We tested two such theoretical ideas using Langevin Dynamics simulations that include realistic field lines. 
The next logical modeling step would be to add hydrodynamic interactions; previous simulation studies suggested that HI have small effects on molecular conformations at capture  \cite{farahpourChainDeformationTranslocation2013} and on translocation times \cite{gauthierMolecularDynamicsSimulation2008}. 

We first proposed using unbiased ZIFE pulses with a feedback mechanism (necessary since ZIFE pulses cannot capture polymers). Although symmetric pulses can preferentially ratchet small captured molecules through the pore, we showed that asymmetric pulses can significantly enhance the device efficiency. Our results demonstrate that temporal asymmetry is sufficient to drive the ratchet process with a symmetric nanopore: asymmetric channels such as periodic cone/sawtooth-shaped channels\cite{mondalRatchetRectificationEffect2016,heidariMechanismRectificationPolymer2020} or spatial asymmetric external potentials (\textit{e.g.} sawtooth-shaped potentials\cite{magnascoForcedThermalRatchets1993,baderDNATransportMicromachined1999,chialvoAsymmetricUnbiasedFluctuations1995,kenwardPolymerDeformationBrownian2008}) are not needed. In essence, the pulsed field reduces the impact of the overlapping distribution functions of DC translocation times.

However, the presence of the feedback loop is likely to be a major issue in practice for several reasons. We thus proposed to add a bias to the ZIFE pulses to build a device that can both capture and translocate short polymer molecules. Since this process does not need a feedback loop, it can be used with a large number of pores in parallel (with a substantial gain in quantities) or in series (with gains in selectivity). In these cases, the mean-field has to be tuned properly to achieve sufficient capture rates while maintaining a useful ratchet effect (the addition of the bias increases the probability that long chains may translocate).

Time varying fields have been proposed previously, but essentially with the goal of increasing sequencing performance and accuracy (see, for example, Refs.~\cite{cherfAutomatedForwardReverse2012} and~\cite{noakesIncreasingAccuracyNanopore2019}). Our goals here are fundamentally different since we are proposing approaches to change concentration ratios in mixed DNA solutions. 

Most polymer ratchet devices that have been proposed use asymmetric channels. What makes our approach unique is that we use a symmetric nanopore system and take advantage of the dynamical asymmetries that exist during translocation (different polymer lengths and conformations across the wall). As we have shown, asymmetric pulses can improve the resulting effects. However, it is also possible that the pore geometry and walls could be designed to further increase the effects reported here. For example, Mondal \textit{et al.} \cite{mondalStochasticResonancePolymer2016} have shown that pore-polymer interactions and conformational entropy can lead to Stochastic Resonance effects during the translocation of flexible polymer chains under an oscillatory driving force. Adding such effects would greatly increase the size of the parameter space and may lead to new ways to manipulate DNA molecules in fluidic systems. 

\begin{acknowledgement}
\begin{sloppypar}
Simulations were performed using the ESPResSo package \cite{weikESPResSoExtensibleSoftware2019} on the computer clusters provided by Compute Canada (https://www.computecanada.ca). GWS acknowledges the support of both the University of Ottawa and the Natural Sciences and Engineering Research Council of Canada (NSERC), funding reference number RGPIN/046434-2013. LQ acknowledges the support of the Mitacs Globalink program and the University of Ottawa. CH and KS are partially funded by Deutsche Forschungsgemeinschaft (DFG, German Research Foundation) under Germany's Excellence Strategy – EXC 2075 – 390740016.
\end{sloppypar}
\end{acknowledgement}

\begin{suppinfo}
\section{Polymer Capture}
In order to fully understand and optimize pulsed field protocols to ratchet semiflexible chains, we must examine polymer capture since this does impact translocation dynamics. We start each capture simulation by placing the center of mass of a polymer with a random equilibrium conformation (radius of gyration $R_{go}$) at a radial distance $r_o \! = \! \lambda_e \! \approx \! 5 \! - \! 10\,R_{go}$ right above the nanopore; the capture is complete when one end monomer has reached the channel's mid-point.

During capture, the electric forces drive molecules to the nanopore, and their conformations deform and orient in response to both the converging field lines and the presence of the wall. This can be visualized by following the mean square radius of gyration $R_g^2$ as the "doomed" chain end approaches the pore -- Fig.~\ref{fig:conformation}a. We note that $R_g^2$ doubles when the end monomer moves from $\approx 8 \,R_{go}$ to $\approx 5\, R_{go}$. However, the conformation then compresses as the chain gets closer to the nanopore. Since the end monomer needs time to find the pore entrance (hence the cloud of data points around $[r=2R_{go},R_g=R_{go}]$), the rest of the chain catches up, further compression takes place, and eventually we even get $R_g\!<\!R_{go}$. Our results are consistent with those of \cite{vollmerTranslocationNonequilibriumProcess2016}.

\begin{figure}[htbp]
\begin{center}
\includegraphics[scale=1.2]{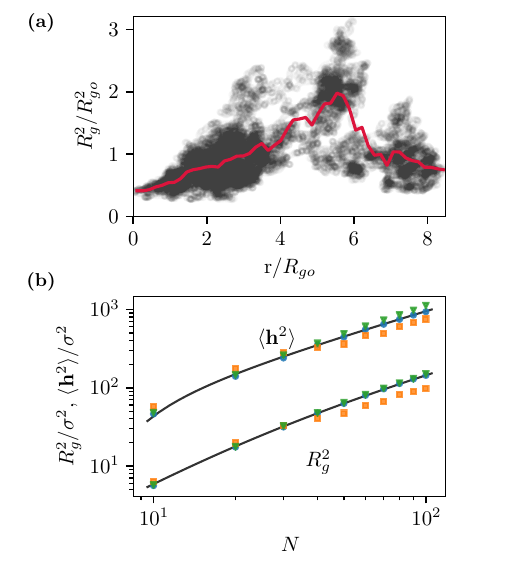}
\end{center}
\caption{(a) Polymer mean square radius of gyration $R_g^2$ \textit{vs} distance $r$ between the pore entrance and the end-monomer that will be captured. Both axes are rescaled by the polymer's equilibrium radius of gyration $R_{go}=8.9\, \sigma$. The grey cycles give periodically sampled simulation data obtained for the capture of a single $N \! = \! 60$ chain with a field intensity $\lambda_e \! = \! 67.5\,\sigma$; the solid line is the binned average. (b) $R_g^2$ and mean squared end-to-end distance $\langle \bf{h}^2 \rangle$ \textit{vs} chain length $N$ (ensemble size $\Omega=1000$). The field intensity is $\lambda_e(N)= \frac{N}{60} \times 67.5 \,\sigma$.  Data shown for polymers in free solution    (\textcolor{NavyBlue}{\Large{$\bullet$}}), after being captured by the pore (\textcolor{Orange}{$\blacksquare$}) and grafted on the wall (\textcolor{OliveGreen}{\large{$\blacktriangledown$}}). The solid lines are the theoretical predictions of the Kratky-Porod equation \cite{teraokaPolymerSolutionsIntroduction2010} for free semi-flexible chains.} 
\label{fig:conformation}
\end{figure}

The semi-flexible nature of the chain impacts this process. Figure~\ref{fig:conformation}b shows the $N$-dependence of both $R_g^2$ and the mean square end-to-end distance $\langle \bf{h}^2 \rangle$ of free chains and of polymers immediately after capture. Since simulation studies often start chains in a relaxed state with $\ge 1$ monomers already engaged in the channel, we have also added data corresponding to the relaxed state of a chain grafted on the wall. Up to $N \! \approx \! 40$, both captured and grafted chains are slightly more extended than free chains; since these molecules are close to the rod-like limit, they orient \cite{qiaoCaptureRodlikeMolecules2020} during capture which helps the end monomer find the pore. But beyond that point, the data sets diverge, with captured chains being more compact (in agreement with Fig.~\ref{fig:conformation}a) and grafted chains being slightly more extended. Theory \cite{ikonenUnifyingModelDriven2012,sarabadaniDrivenTranslocationSemiflexible2017} indicates that initial conformations affect translocation times; indeed, for the range of chain lengths and field intensities used here, we obtain $\overline{\tau}=(1.10\pm 0.03)\times N^{1.21\pm 0.01}$ when we start with captured conformations, but $\overline{\tau}=(0.95\pm0.08)\times N^{1.28\pm0.02}$ for relaxed chains (data not shown). The field outside the pore also plays a role since we find $\overline{\tau}=(0.89\pm 0.10)\times N^{1.31\pm0.02}$ when relaxed conformations are driven with the force applied only in the nanopore, a standard set-up in simulations.
\end{suppinfo}

\bibliography{Refs_final}

\end{document}